\documentstyle[11pt,newpasp,twoside,psfig,epsf]{article}
\def\edcomment#1{\iffalse\marginpar{\raggedright\sl#1\/}\else\relax\fi}
\reversemarginpar

\begin{document}
\title{TeV Gamma-Ray Absorption and the Intergalactic Infrared 
Background}

\author{F.W. Stecker}
\affil{Laboratory for High Energy Astrophysics, NASA Goddard Space Flight 
Center, Greenbelt, MD 20771, USA}

\begin{abstract}

In this paper, I will take a synoptic approach to determining the intergalactic
infrared radiation field (IIRF). This approach draws on both the multi-TeV
$\gamma$-ray observations and the infrared background observations and relates
them {\it via} the semi-empirical modelling method of Malkan \& Stecker.
I discuss the evidence for an intergalactic infrared background obtained
by an analysis of the {\it HEGRA} observations of the high energy $\gamma$-ray
spectrum of Mrk 501 and from constraints from Mrk 421 deduced from the Whipple 
air Cherenkov telescope results. I will show that this evidence is in accord
with the predictions made by Malkan \& Stecker (1998) for the intergalactic
infrared spectral energy distribution produced by galaxies. The Malkan-Stecker
predictions are also in excellent agreement with mid- and far infrared galaxy 
counts. However, there may be potential problems relating these predictions 
with the results of the analysis of {\it COBE-DIRBE} far infrared data. 
The $\gamma$-ray and {\it COBE-DIRBE} observations may also need to be
reconciled. I will discuss possible ways to resolve this situation including 
a partial nullification of the $\gamma$-ray absorption process which
can hypothetically occur if Lorentz invariance is broken.

\end{abstract}

\section{Introduction}

Shortly after the discovery that blazars can be strong sources of high
energy $\gamma$-rays, we suggested that
very high energy $\gamma$-ray beams from blazars can be used to 
measure the intergalactic infrared radiation field, since 
pair-production interactions of $\gamma$-rays with intergalactic IR photons 
will attenuate the high-energy ends of blazar spectra (Stecker, De Jager \&
Salamon 1992).
In recent years, this concept has been used successfully to place upper limits 
on the the intergalactic IR field (IIRF) (Stecker \& De Jager 1993; 
Dwek \& Slavin 1994; Stecker \& De Jager 1997; Stanev \& Franceschini 1997;
Biller {\it et al.} 1998).
Determining the IIRF, in turn, allows us to 
model the evolution of the galaxies which produce it. 
As energy thresholds are lowered 
in both existing and planned ground-based
air Cherenkov light detectors, cutoffs in the $\gamma$-ray 
spectra of 
more distant blazars are expected, owing to extinction by the IIRF. These
can be used to explore the redshift dependence of the 
IIRF.

On the other hand, by modelling the intergalactic infrared and optical 
radiation fields as a function of redshift, one can calculate the expected 
$\gamma$-ray opacity as a function of redshift and comparison of predicted
absorption with the data can yield important new information about the 
evolution of the IIRF (Stecker \& De Jager 1998; Salamon \& Stecker 1998). 

Exploring the IIRF with high energy $\gamma$-rays has both advantages and
disadvantages over direct infrared observations. An important advantage is that
since the absorption process only acts effectively in intergalactic space over
large distances, the $\gamma$-ray approach does not suffer from
problems of subtraction of foreground emission from both zodiacal light and
interstellar dust in the Galaxy. As we have seen at this meeting, these
can be formidable sources of uncertainty in determining the IIRF. Also,
high energy $\gamma$-ray observations of blazar and $\gamma$-ray burst
spectra at high redshifts can be used to probe the past evolution of the IIRF
whereas direct infrared observations can only tell us about the present IIRF. 

On the other hand, the disadvantage of the $\gamma$-ray approach is the
uncertainty in the unabsorbed spectrum of the source. This uncertainty
is somewhat ameliorated by the strong energy dependence of the absorption
effect itself.
Therefore, it is my contention that
using both direct infrared and high energy $\gamma$-ray 
observations in a ``synoptic'' approach can yield the most information
about the IIRF.

\section{The Opacity of Intergalactic Space Owing to the IIRF}

The formulae relevant to absorption calculations involving pair-production 
are given and discussed in Stecker, De Jager \& Salamon (1992), where we 
derived the absorption
formulae with cosmological and redshift effects included.
For $\gamma$-rays in the TeV energy range, the pair-production cross section 
is maximized when the soft photon energy is in the infrared range:
\begin{equation} 
\lambda (E_{\gamma}) \simeq \lambda_{e}{E_{\gamma}\over{4m_{e}c^{2}}} =
1.24E_{\gamma,TeV} \; \; \mu m 
\end{equation}
where $\lambda_{e} = h/(m_{e}c)$ 
is the Compton wavelength of the electron.
For 15 TeV $\gamma$-rays, absorption will occur primarily by interactions
with mid-infrared photons having a
wavelength $\sim$ 20$\mu$m. (Pair-production interactions actually
take place with photons over a range of wavelengths around the optimal value as
determined by the energy dependence of the cross section; see eq. (11)).) 
If the emission spectrum of
an extragalactic source extends beyond 20 TeV, then the extragalactic
infrared field should cut off the {\it observed} spectrum between $\sim
20$ GeV and $\sim 20$ TeV, depending on the redshift of the source (Stecker
\& de Jager 1998; Salamon \& Stecker 1998).

\section{Absorption of Gamma-Rays at Low Redshifts}

Stecker \& De Jager (1998) (hereafter SD98) have calculated the 
absorption coefficient of intergalactic
space using a new, empirically based calculation
of the spectral energy distribution (SED) of intergalactic low energy 
photons by Malkan \& Stecker (1998) (hereafter MS98) 
obtained by integrating luminosity dependent infrared spectra of galaxies
over their luminosity and redshift distributions.
After giving their results on the $\gamma$-ray optical depth as a function of energy 
and redshift out to a redshift of 0.3, Stecker \& De Jager (1998) (SD98) 
applied their calculations by
comparing their results with the spectral data on Mrk 421 (McEnery {\it
et al.} 1997) and 
spectral data on Mrk 501 (Aharonian, {\it et al.} 1997). 

SD98 made the reasonable simplifying assumption 
that the IIRF is basically in
place at a redshifts $<$ 0.3, having been produced primarily at higher
redshifts (Madau 1995; Salamon \& Stecker 1998). 
Therefore SD98 limited their calculations to $z<0.3$. 

\begin{figure}
\centerline{\psfig{file=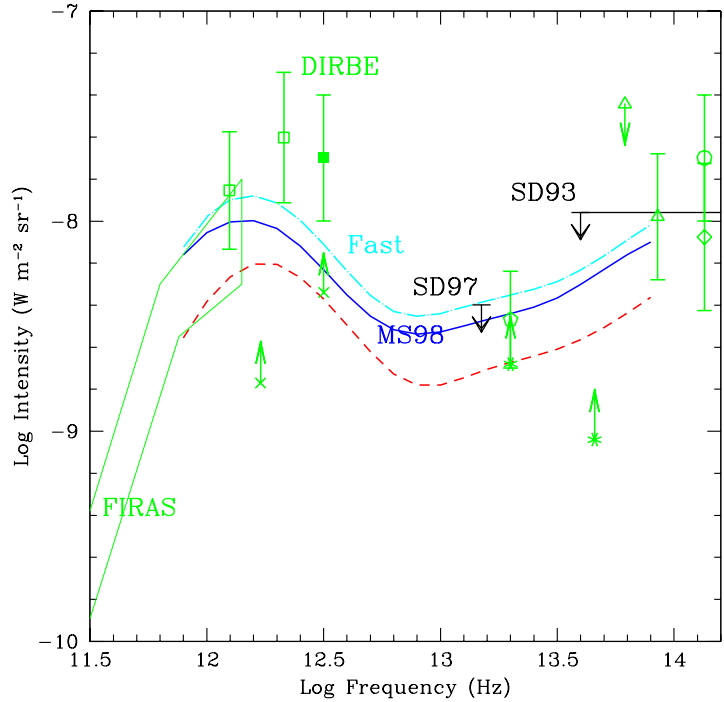,width=10.0truecm}}
\caption{The spectral energy distribution (SED) of the extragalactic 
infrared radiation calculated
by Malkan \& Stecker (1998, 2000) The dashed line (lower IIRF curve) and the 
solid line 
(middle IIRF curve) correspond to the middle and upper curves calculated by 
Malkan \& Stecker (1998) with redshift-evolution assumptions as 
described in the text. The dot-dashed line (Malkan \& Stecker 2000)
corresponds to a ``fast evolution'' case with $Q=4.1$ and $z_{flat}=1.3$. 
Representative data are also shown {\it with 2$\sigma$ error bars} as 
follows. From left to right, {\it FIRAS} polygon: Fixsen {\it et al.} 1998, 
open squares: Hauser {\it et al.} (1998), solid square: 
Lagache {\it et al.} (2000), pentagon: 
Altieri {\it et al.} (1999), open triangles:
Dwek and Arendt (1998), open circle: Gorjian, Wright \& Chary (2000), 
diamond: Totani, {\it et al.} (2000) (see also Pozzetti {\it et al.} (2000). 
The lower limits with crosses are from {\it ISOPHOT} (Puget {\it et al.} 1999) 
and the lower limits with asterisks are from {\it ISOCAM} (Elbaz {\it et al.} 
(1999). The upper limits are from Stecker \& de Jager (1993)(SD93) and Stecker \& de Jager (1997)(SD97).}
\label{irdata}

\end{figure}

SD98 assumed for the IIRF, two of the spectral energy distributions (SEDs) 
given in MS98 (shown in Figure 1).
The lower curve in Figure 1 (adapted from MS98) assumes luminosity
evolution proportional to $(1+z)^{3.1}$ 
out to $z=1$, whereas the middle curve assumes such evolution out to $z=2$.
Evolution in stellar emissivity is expected to level off 
at redshifts greater than $\sim 1.5$ (Steidel 1999; Hopkins, Connolley \&
Szalay 2000) 
Using these two SEDs for the IIRF, SD98
obtained parametric expressions for the optical depth $\tau(E_{\gamma},z)$ 
for $z<0.3$, taking a Hubble constant of $H_o=65$ km s$^{-1}$Mpc$^{-1}$  
(Gratton 1997).

The double-peaked form of the SED of the IIRF requires
a third order polynomial to approximate the opacity $\tau$ 
in a parametric form. SD98 give the following approximation:

\begin{equation}
log_{10}[\tau(E_{\rm TeV},z)]\simeq\sum_{i=0}^3a_i(z)(\log_{10}E_{\rm TeV})^i
\;\;{\rm for}\;\;1.0<E_{\rm TeV}<50,
\end{equation}
where the z-dependent coefficients are given by
\begin{equation}
a_i(z)=\sum_{j=0}^2a_{ij}(\log_{10}{z})^{j}.
\end{equation}
Table 1 gives the numerical values for 
$a_{ij}$, with $i=0,1,2,3$, and $j=0,1,2$. The numbers before the
brackets are obtained using the lower IIRF SED shown in Figure 1; The
numbers in the brackets are obtained using the middle IIRF SED.
Equation (2) approximates $\tau(E,z)$ to within 10\% for all values of z and E 
considered. Figure 2 shows the results obtained by SD98 for 
$\tau(E_{\gamma},z)$.

\vspace{1em}

\begin{tabular}{|c|r|r|r|r|}
\multicolumn{5}{c}{\bf Table 1: Polynomial coefficients $a_{ij}$}\\
\hline
$j$ & $a_{0j}$ & $a_{1j}$ & $a_{2j}$ & $a_{3j}$ \\ \hline
0&1.11(1.46) &-0.26(~0.10) &1.17(0.42) &-0.24(~0.07)\\
1&1.15(1.46) &-1.24(-1.03) &2.28(1.66) &-0.88(-0.56)\\
2&0.00(0.15) &-0.41(-0.35) &0.78(0.58) &-0.31(-0.20)\\ \hline
\end{tabular}

\vspace{1em}

The advantage of using empirical
data to construct the SED of the IIRF, as done in MS98, is 
particularly indicated in the mid-infrared range. In this region of 
the spectrum, galaxy observations indicate more flux from warm 
dust in galaxies than that
taken account of in more theoretically oriented models 
({\it e,g,} MacMinn \& Primack (1996)). As a consequence, the 
mid-infrared ``valley'' between the cold 
dust peak in the far-infrared and cool star peak in the near IR is 
filled in more in the MS98 results and is
not as pronounced as in previously derived models of the IR background SED.
As can be seen in Figure 1, the background SED predicted in MS98 is in 
excellent agreement with the results
obtained from ultradeep {\it ISOCAM} galaxy count observations at 15 $\mu$m
(Altieri {\it et al.} 1999).

\begin{figure}
\vspace{1.0truecm}
\centerline{\psfig{file=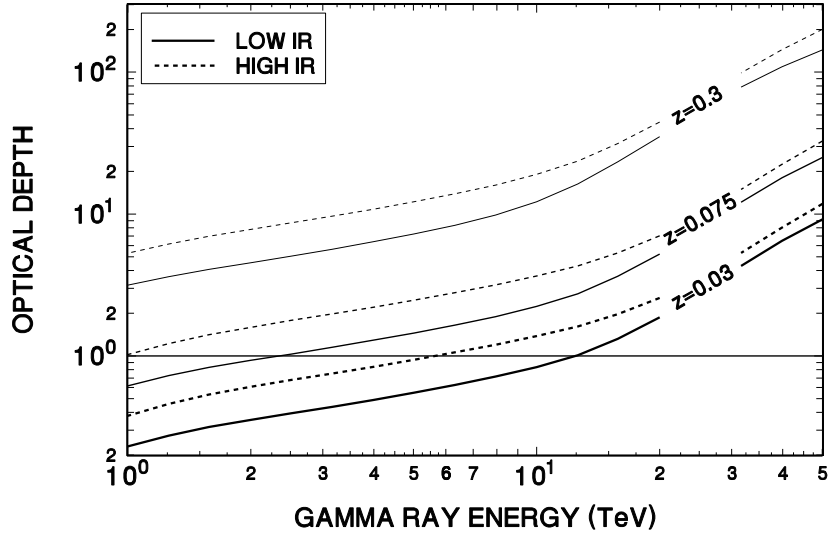,width=13.0truecm}}
\vspace{-8.0truecm}
\caption{
Optical depth versus energy for $\gamma$-rays originating at various redshifts
obtained using the SEDs corresponding to the lower IIRF (solid lines) and 
middle IIRF (dashed lines) levels shown in Fig. 1 (from SD98). Note that the
line styles in this Figure are the {\it reverse} of those in Figure 1.}
\label{lowztau}
\end{figure}
\vskip 2.0truecm

The SD98 calculations predict 
that intergalactic absorption should only 
slightly steepen the spectra of Mrk 421 and Mrk 501 below $\sim$ 7 TeV,
which is consistent with the data in the published literature
The SD98 calculations further predict that
intergalactic absorption should turn over the spectra of these sources 
at energies greater than $\sim$ 15 TeV.

\section{The Multi-TeV Spectrum of Mrk 501 and its Interpretation}

The {\it HEGRA} group has observed the BL Lac object Mrk 501 in the flaring 
phase, obtaining an energy spectrum for this source up  
to an energy of 24 TeV (Aharonian {\it et al.} 1999).
The Mrk 501 spectrum obtained by the {\it HEGRA} group are well fitted
by a source spectrum power-law of spectral index $\sim 2$ steepened at
in the multi-TeV energy range by intergalactic absorption, with the 
optical depth calculated by SD98 (Konopelko, {\it et al.} 1999). Figure 3 
clearly shows this.

\begin{figure}
\vskip -3.0truecm
\epsfysize=8.0truein
\centerline{\epsfbox{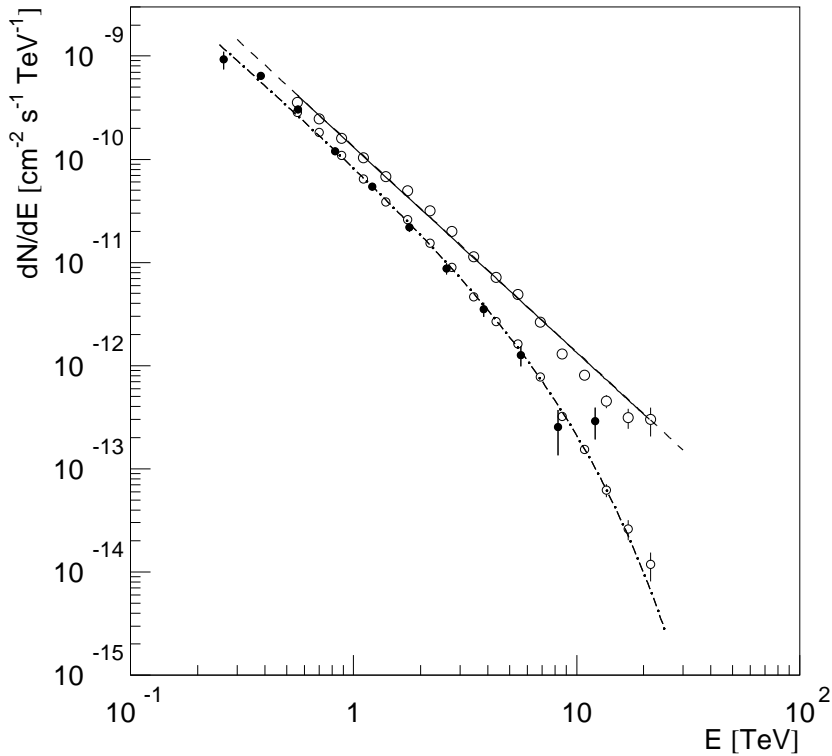}}
\vskip -5.0truecm
\caption{The {\it HEGRA} and Whipple telescope data on Mrk 501 
in the flaring state is indicated by the lower line. 
The upper line and points show
the intrinsic spectrum of the source with the effect of extragalactic
absorption removed. The absorption is calculated using the middle line
IIRF spectrum shown in Figure 1 (see also Figure 2) 
(From Konopelko {\it et al.} 1999).}
\vskip 1.0truecm
\label{hegra}
\end{figure} 

\section{Reconciling the Infrared and Gamma-Ray Observations}

Malkan \& Stecker (2000) have used their empirically based model (MS98) 
to predict infrared luminosity
functions and deep infrared galaxy counts at various wavelengths.
They have also examined their predictions for the IIRF for comparison with the
subsequent determinations from the {\it COBE-DIRBE} data analysis.
Using the formalism of luminosity evolution proportional to $(1+z)^Q$ out to
a redshift of $z_{flat}$ and constant (no evolution) for $z_{flat} < z < 
z_{max} = 4$,
they find that a comparison of their predictions with current {\it ISO} 
galaxy counts at 15 and 175$\mu$m favor their ``Baseline Model" with $Q = 3.1$ 
and $z_{flat} = 2$ (the middle curve in Figure 1). The $\gamma$-ray 
limits (SD97) also favor $Q \sim 3$.

On the other hand, the {\it COBE-DIRBE} 
far infrared determinations seem to favor a stronger evolution with 
$Q > 4$ up to $z_{flat} = 1$. For example, the upper curve in Figure 1
(Malkan \& Stecker 2000) assumes $Q = 4.1$ and $z_{flat} = 1.3$.
This curve gives a flux in the mid-infrared which is a factor of $\sim$1.8
above the ``Baseline Model'' and would imply a corresponding increase in
the opacity of the Universe to multi-TeV $\gamma$-rays. (Thus, one should
multiply the appropriate bracketed numbers in Table 1 by a factor of 1.8)
Therefore, $\tau$(15 TeV) would increase from $\sim 2$ to $\sim 4$ in the
case of Mrk 501 and Mrk 421 ($z \simeq 0.03$).

This {\it prima facie} conflict can be resolved in two ways: either
(a) the {\it COBE-DIRBE} far-infrared estimates may suffer from 
undersubtraction of foreground emission and therefore are too high, or (b) 
the {\it ISOPHOT} galaxy 
counts may be missing a significant fraction of sources. In this later case,
one may also have require that the $\gamma$-ray results are wrong in that
the {\it HEGRA} energy determinations have been overestimated, mimicing
the effect of absorption which would be produced by a lower IIRF. Another
possibility is one involving new physics, {\it viz.} that Lorentz invariance
may be broken, allowing the Universe to be transparent to multi-TeV photons
(Coleman \& Glashow 1999; Kifune 1999). This ``new physics''scenario presents 
problems in that the Mrk 501 spectrum does exhibit exactly the characteristics
expected for high-energy $\gamma$-ray absorption from pair-production 
(Konopelko {\it et al.} 1999). However, we will discuss this possibility
more quantitatively in the next section (Glashow \& Stecker 2000).  

Possibility (a) finds support in the independent analysis of the {\it COBE}
data by Lagache {\t et al.} (1999) who obtain a flux at 140 $\mu$m 
which is only 60\% of the flux obtained by 
Hauser {\it et al.} (1998) shown in Figure 1. 
Lagache {\it et al.} (2000) also obtained a smaller flux at 240 $\mu$m. 
In this regard, one should also note that the results reported by 
Hauser {\it et al.} (1998) were at the 4$\sigma$ level. 

If the far-infrared galaxy counts are incomplete (possibility (b)), 
this would imply stronger evolution in the far-infrared emission of 
galaxies than in the mid-infrared.
Although the MS98 model already includes some differential 
evolution of this type, based on the data of Spinoglio {\it et al.} (1995),
it is conceivable that starburst galaxies at redshifts at $\sim 1$ might
produce an even higher ratio of $\sim 60\mu$m to $\sim 7\mu$m rest-frame 
fluxes than their present-day counterparts. Radiation at these widely 
separated wavelengths  
is known to be emitted by quite different dust grains which could have 
different evolutionary development, particularly for ULIRGs (ultraluminous 
infrared galaxies) and AGN (active galactic nuclei). One should note that 
{\it this possibility can make the $\gamma$-ray and infrared data compatible}, 
since absorption of $\sim$ 15 TeV 
$\gamma$-rays is caused by interactions with mid-infrared ($\sim$ 20$\mu$m)
photons and not 140$\mu$m far-infrared photons.

\section{Breaking Lorentz Invariance}

With the idea of spontaneous symmetry breaking in particle physics came the
suggestion that Lorentz invariance (LI) might be weakly broken at high energies
(Sato \& Tati 1972). Although no real quantum theory of gravity exists, it 
was suggested that LI might be broken as a consequence of such a theory
(Amelino-Camilia {\it et al.} 1998). A simpler formulation
for breaking LI by a small first order perturbation in the electromagnetic 
Lagrangian which leads to a renormalizable treatment has been given by
Coleman \& Glashow (1999). The breaking of LI at high energies is one way
of avoiding the predicted, but unseen, ``GZK'' cutoff in the ultrahigh energy 
cosmic-ray spectrum owing to photomeson interactions with 2.7K cosmic
background photons (Greisen 1966; Zatsepin 
\& Kuz'min 1966), which produces an effective absorption mean-free-path
for ultrahigh energy cosmic rays in intergalactic space of $< 100$ Mpc
(Stecker 1968).

It has recently been suggested that LI breaking could also
make the universe transparent to high energy $\gamma$-rays (Kifune 1999).
Such LI breaking implies a perferred frame of reference in the Universe which
would naturally be associated with a rest frame for the 2.7K cosmic 
background radiation. 
We follow here the formalism proposed by Coleman \& Glashow (1999) for
LI breaking. Within this scenario, the maximum attainable velocity of an
electron, $c_{e} \ne c_{\gamma}$, the velocity of the photon.
Let us consider the case where $c_{e} > c_{\gamma}$ and we define

\begin{equation}
c_{e} \equiv c_{\gamma}(1 + \delta) ~ , ~ ~~~ \delta \ll 1  
\end{equation}

In this case, electrons above an energy $E_{max} = \gamma_{max}m_{e}$ will be
superluminal. 
They will radiate Cherenkov light if their velocity is large
enough, {\it i.e.}, if $\beta > (c_{\gamma}/c_{e})$
which implies that $\gamma_{e}^2 > [1 - (c_{\gamma}/c_{e})^2]^{-1}$. 
This determines the maximum electron energy above which electrons would rapidly
lose energy by Cherenkov radiation:

\begin{equation}
E_{max} = m_{e}(2\delta)^{-1/2}
\end{equation}

\noindent (In this section, we adopt the standard paticle physics convention 
$c = 1$.) 
Since electrons are seen in the cosmic radiation up to an energy $\sim$ 1 TeV,
this implies that $E_{max} > 1$ TeV which gives an upper limit on $\delta$
of $1.3 \times 10^{-13}$. 

If LI is broken so that $\delta > 0$, the threshold energy for the pair 
production process is altered because the square of the four-momentum becomes

\begin{equation}
2\epsilon E_{\gamma}(1 - \cos \theta) - 2E_{\gamma}^2\delta = 4\gamma^2m_{e}^2 >4 m_{e}^2
\end{equation}

\noindent where $\epsilon$ is the energy of the low energy (infrared) photon and $\theta$
is the angle between the two photons. The second term on the left-hand-side
comes from the fact that $c_{\gamma} =  
\partial E_{\gamma}/\partial p_{\gamma}$.

For head-on collisions ($\cos \theta = -1$) the minimum low energy photon
energy for pair production becomes 

\begin{equation}
\epsilon_{min} = m_{e}^2/E_{\gamma} +  (E_{\gamma}\delta)/2
\end{equation}

It follows that the condition for a significant increase in the energy
threshold for pair production is 

\begin{equation}
(E_{\gamma}\delta)/2 \ge m_{e}^2/E_{\gamma}  
\end{equation}

\noindent or, equivalently, $\delta$ must be greater than 
$2m_{e}^2/E_{\gamma}^2$.

Thus, for a significant decrease in the optical depth to Mrk 501 for 
$E_{\gamma} = 15$ TeV, we must have $\delta \ge 2.4 \times 10^{-15}$.

The effect of breaking LI is to exclude photons of energy below 
$\epsilon_{min}$ from pair producing, thus reducing the number of target
photons. In this way, one can exclude the high flux of far-infrared photons
implied by the {\it COBE-DIRBE} analysis from participating in the absorption 
process while still allowing mid-infrared photons to produce an absorption 
feature.

One can rule out this hypothesis is if one observes electrons with energies 
$\gg$ 1 TeV. This would reduce the upper limit on $\delta$ to the point where 
there will be no significant reduction in absorption by pair-production
interactions with infrared photons.  
   
On the other hand, if one observes $\gamma$-rays above 100 TeV from an
extragalactic source, this would be strong evidence for LI breaking. This is
because the very large density ($\sim$ 400 cm$^{-3}$) of 3K cosmic microwave
photons would otherwise absorb $> 100$ TeV $\gamma$-rays within a distance of
$\sim$ 10 kpc (see eq. (1)).

Finally, we note that even if absorption is reduced in Mrk 501, one can still
have absorption in sources at higher redshifts owing to interactions with
higher energy photons (see next section).
 
\section{Absorption of Gamma-Rays at High Redshifts}

We now discuss the absorption of 10 to
500 GeV $\gamma$-rays at high redshifts. 
In order to calculate such high-redshift absorption properly, it is
necessary to determine the spectral distribution of the intergalactic low 
energy photon background radiation as a function of redshift as realistically 
as possible out to frequncies beyond the Lyman limit. This calculation,
in turn, requires observationally based information on the evolution of the 
spectral energy distributions (SEDs) of IR through UV starlight from galaxies,
particularly at high redshifts. 

Conversely, observations of high-energy cutoffs in the
$\gamma$-ray spectra of blazars as a function of redshift, which may enable one to 
separate out intergalactic absorption from redshift-independent cutoff 
effects, could add to our knowledge of galaxy formation and early galaxy 
evolution. In this regard, it should be noted that the study of blazar spectra in the 10 to 300 GeV range is one of the primary goals of a next generation
space-based $\gamma$-ray  telescope {\it GLAST} 
(Gamma-ray Large Area Space Telescope)
(Bloom 1996) 
as well as {\it VERITAS} and 
other future ground based $\gamma$-ray telescopes. 

Salamon and Stecker (1998) (hereafter SS98) have calculated 
the $\gamma$-ray opacity as a function of both energy and redshift
for redshifts as high as 3 by taking account of the evolution of both the
SED and emissivity of galaxies with redshift. 
In order to accomplish this, 
they adopted the recent analysis of Fall {\it et al.} (1996) and 
also included the effects of metallicity evolution on galactic SEDs.
Their results indicate that the extragalactic
$\gamma$-ray background spectrum from blazars should steepen significantly 
above 20 GeV, owing to extragalactic absorption. Future observations of
such a steepening would thus provide a test of the blazar origin
hypothesis for the $\gamma$-ray background radiation. 
      
\subsection{Redshift Dependence of the Intergalactic Low Energy SED}

Pei and Fall (1995) have devised a method for calculating stellar 
emissivity which bypasses the uncertainties associated with estimates of 
poorly defined luminosity distributions of evolving galaxies.
The core idea of their 
approach is to relate the star formation rate 
directly to the evolution of the neutral gas density in damped
Lyman $\alpha$ systems, and then to use stellar population synthesis models to
estimate the mean co-moving stellar emissivity ${\cal E}_{\nu}(z)$
of the universe as a function of frequency $\nu$ and
redshift $z$.
The SS98 calculation of stellar emissivity closely follows this 
elegant analysis, with minor modifications.

Damped Lyman $\alpha$ systems are high-redshift clouds of gas whose neutral
hydrogen surface density is large enough ($>2\times 10^{20}$ cm$^{-2}$)
to generate saturated Lyman $\alpha$
absorption lines 
in the spectra of background quasars that 
happen to lie along and
behind  common lines of sight to these clouds.  
These gas systems are believed
to be either precursors to galaxies or young galaxies themselves, 
since their neutral hydrogen (HI)
surface densities are comparable to those of spiral galaxies today, and their
co-moving number densities are consistent with those of present-day galaxies
(Worthy 1986; Pei \& Fall 1995).  It is in these systems that initial
star formation presumably took place, so there is a relationship between
the mass content of stars and of gas in these clouds; if there is no infall
or outflow of gas in these systems, the systems are ``closed'', so that the
formation of stars must be accompanied by a reduction in the neutral gas
content.  Such a variation in the HI surface densities of
Lyman $\alpha$ systems with redshift is seen, and is used by Pei and Fall (1995)
to estimate the mean cosmological rate of star formation back to 
redshifts as large as $z=5$.

Following Fall, Charlot \& Pei (1996),
SS98 used the Bruzual-Charlot model (Charlot \& Bruzual 1991; Bruzual \&
Charlot 1993) corresponding to a Salpeter stellar initial mass function,
$\phi(M)\,dM\propto M^{-2.35}\,dM$, where $0.1M_{\odot}<M<125M_{\odot}$.
The mean co-moving emissivity ${\cal E}_{\nu}(t)$
was then obtained by convolving over time $t$ 
the specific luminosity 
with the mean co-moving mass rate of star formation.
SS98 also obtained metallicity correction
factors for stellar radiation at various wavelengths. Increased metallicity 
gives a redder population spectrum (Worthy 1994; Bertelli, {\it et al.} 
1994).

SS98 calculated stellar
emissivity as a function of redshift at 0.28 $\mu$m, 0.44 $\mu$m,
and 1.00 $\mu$m, both with and without a metallicity correction.
Their results agree well with the
emissivity obtained by the Canada-French
Redshift Survey (Lilly, {\it et al} 1996) over the redshift range of the 
observations
($z \le 1$).

The stellar emissivity in the universe is found to peak
at $ 1 \le z \le 2$, dropping off steeply at lower reshifts and more slowly
at higher redshifts. Indeed,
Madau, Pozzetti \& Dickinson (1998) have used observational data from the 
Hubble Deep Field to show that metal production has a similar redshift 
distribution, such production being a direct measure of the star formation 
rate.

The co-moving radiation energy density $u_{\nu}(z)$ 
is the time integral of the co-moving emissivity ${\cal E}_{\nu}(z)$,
\begin{equation} \label{B}
u_{\nu}(z)=
\int_{z}^{z_{\rm max}}dz^{\prime}\,{\cal E}_{\nu^{\prime}}(z^{\prime})
\frac{dt}{dz}(z^{\prime})e^{-\tau_{\rm eff}(\nu,z,z^{\prime})},
\end{equation}
where $\nu^{\prime}=\nu(1+z^{\prime})/(1+z)$ and $z_{\rm max}$ is the
redshift corresponding to initial galaxy formation.
The extinction term $e^{-\tau_{\rm eff}}$ 
accounts for the absorption of ionizing photons by the clumpy
intergalactic medium (IGM) that lies between the source and observer.
Although the IGM is effectively transparent to non-ionizing photons,
the absorption of photons by HI,
HeI and HeII can be considerable (Madau 1995).

\subsection{The Gamma-Ray Opacity at High Redshifts}

SS98 evaluated the co-moving energy density $u_{\nu}(z)$ using
equation (9) above. They then calculated 
the optical depth for $\gamma$-rays owing to electron-positron pair production 
interactions with photons of the stellar radiation
background. This can be determined from the expression (Stecker, De Jager \&
Salamon 1992)

\begin{equation} \label{G}
\tau(E_{0},z_{e})=c\int_{0}^{z_{e}}dz\,\frac{dt}{dz}\int_{0}^{2}
dx\,\frac{x}{2}\int_{0}^{\infty}d\nu\,(1+z)^{3}\left[\frac{u_{\nu}(z)}
{h\nu}\right]\sigma_{\gamma\gamma}(s)
\end{equation}
where $s=2E_{0}h\nu x(1+z)$,
$E_{0}$ is the observed $\gamma$-ray energy at redshift zero, 
$\nu$ is the frequency at redshift $z$,
$z_{e}$ is the redshift of
the $\gamma$-ray source, $x=(1-\cos\theta)$, $\theta$ being the angle between
the $\gamma$-ray and the soft background photon, $h$ is Planck's constant
($\epsilon = h\nu$), and
the pair production cross section $\sigma_{\gamma\gamma}$ is zero for
center-of-mass energy $\sqrt{s} < 2m_{e}c^{2}$, $m_{e}$ being the electron
mass.  Above this threshold, 
\begin{equation} \label{H}
\sigma_{\gamma\gamma}(s)=\frac{3}{16}\sigma_{\rm T}(1-\beta^{2})
\left[ 2\beta(\beta^{2}-2)+(3-\beta^{4})\ln\left(\frac{1+\beta}{1-\beta}
\right)\right],
\end{equation}
where $\beta=(1-4m_{e}^{2}c^{4}/s)^{1/2}$.

Figure 4, based on the results of SS98, shows the predicted
critical energy for a 1/e absorption by pair production versus redshift
obtained for the cases of correction for metallitcity evolution and no
correction. Absorption of $\gamma$-rays of energies below $\sim$15 GeV is
negligible.
~
\begin{figure}[t]
\epsfysize=4.0truein
\centerline{\epsfbox{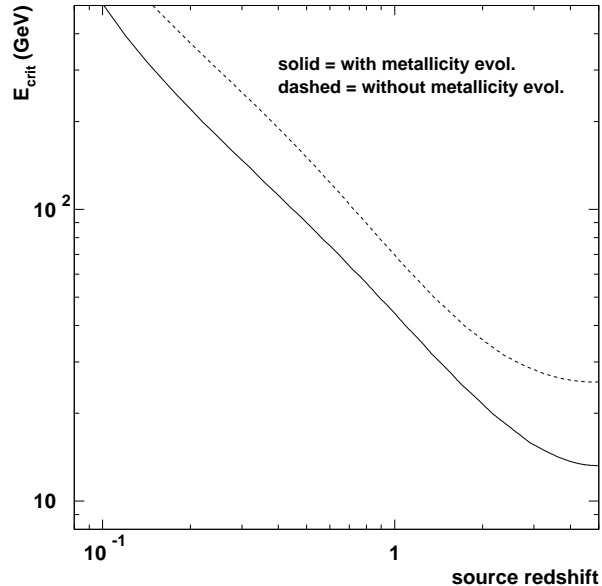}}
\vspace{-1.0truecm}
\caption{The critical energy for $\gamma$-ray absorption above which
the optical depth is predicted to be greater than 1 as a function of
the redshift of the source (obtained from the results of SS98).}
\label{ecritvsz}
\end{figure}

The weak redshift dependence of the opacity at the higher redshifts, 
as shown in Figure 4, indicates that the opacity is not very
sensitive to the initial epoch of galaxy formation, contrary to the 
speculation of MacMinn and Primack (1996). In fact, the uncertainty in the
metallicity correction (see Figure 4) would obscure 
any dependence on $z_{max}$ even further.

\begin{figure}
\centerline{\psfig{file=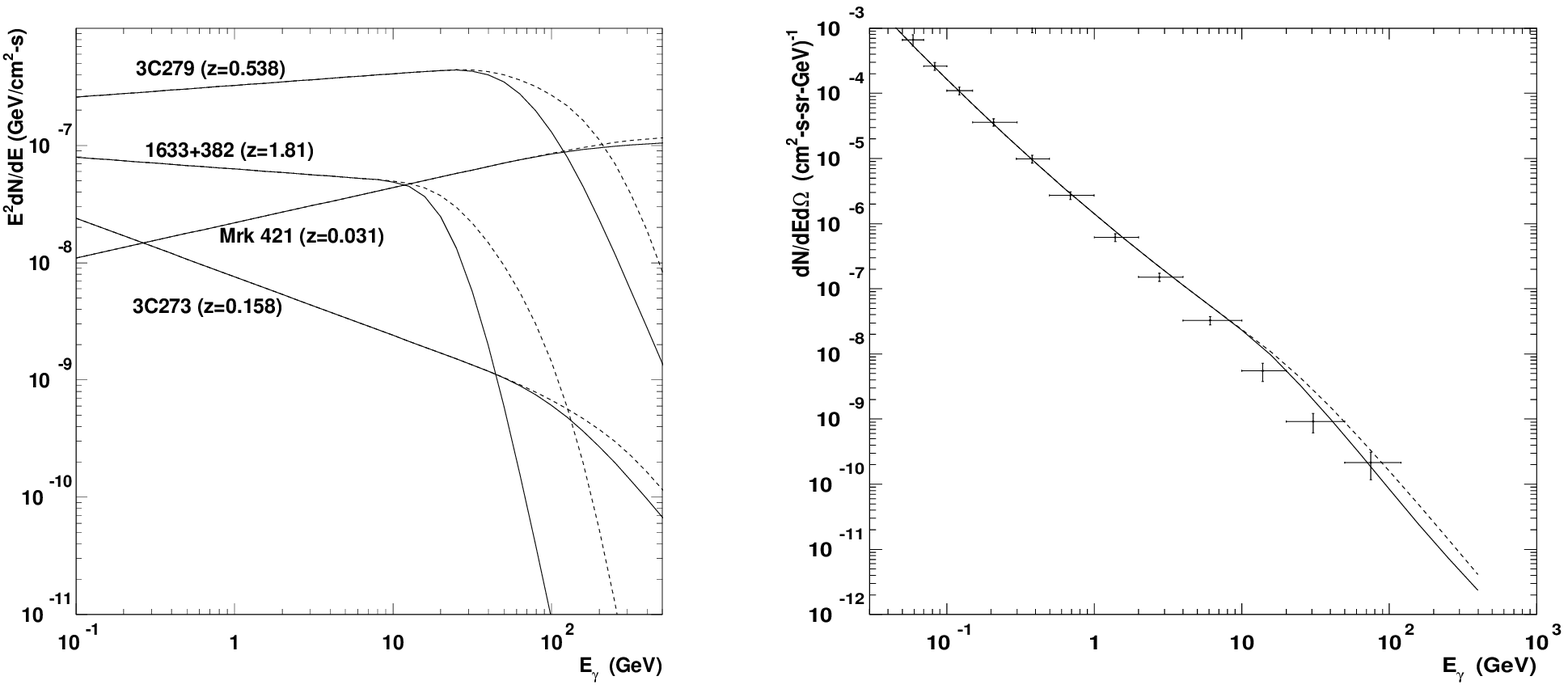,width=12.0truecm}}
\caption{ The left graph shows the effect of intergalactic absorption by 
pair-production on the power-law spectra of
four prominent blazars: 1633+382 ($z=1.81$), 
3C279 ($z=0.54$), 3C273 ($z=0.15$), and Mrk 421 ($z=0.031$); The right
graph shows the
extragalactic $\gamma$-ray background spectrum predicted by the unresolved 
blazar model of
Stecker \& Salamon (1996) with absorption included, calculated for a mean 
{\it EGRET} point-source 
sensitivity of $10^{-7}$ cm$^{-2}$s$^{-1}$, compared with the {\it EGRET}
data on the $\gamma$-ray background (Sreekumar {\it et al.} 1998).  
The solid (dashed) curves are calculated with (without) the metallicity
correction function (from SS98).}
\label{blaz}
\end{figure}

\subsection{The Effect of Absorption on the Spectra of Blazars and the
Gamma-Ray Background}

With the $\gamma$-ray opacity $\tau(E_{0},z)$ calculated out to
$z=3$,
the cutoffs in blazar $\gamma$-ray spectra caused by extragalactic pair 
production interactions with stellar photons can be predicted.
The left graph in Figure 5, from SS98,
shows the effect of the intergalactic radiation
background on a few of the blazars 
observed by {\it EGRET},
{\it viz.}, 1633+382, 3C279, 3C273, and Mrk 421,
assuming that the mean spectral indices obtained for these sources by
{\it EGRET} extrapolate out to higher energies 
attenuated only by intergalactic 
absorption.  (Observed cutoffs in blazar spectra 
may be intrinsic cutoffs in $\gamma$-ray production in
the source, or may be caused
by intrinsic $\gamma$-ray absorption within the source itself.) 

The right hand graph in Figure 5
shows the background spectrum predicted from unresolved blazars
(Stecker \& Salamon 1996; Salamon \& Stecker 1998)
compared with the {\it EGRET} data (Sreekumar, {\it et al.} 1998). 
Note that the 
predicted spectrum steepens 
above 20 GeV, owing to extragalactic absorption by pair-production 
interactions with intergalactic optical and infrared photons, particularly at
high redshifts.

\section{Conclusions}

Studies of the absorption of high energy $\gamma$-rays 
in the spectra of extragalactic objects can be used to probe both the present
intergalactic infrared background and the infrared and optical radiation that
existed in intergalactic space at higher redshifts. The results of such
observations are free of the effects of solar system and galactic foreground
contamination. Such studies are most effective when combined with direct
infrared observations in a synoptic approach.

High energy $\gamma$-ray observations have already provided the best 
constraints on the mid-infrared background and they support a galaxy
evolution scenario where emissivity evolution in the mid-infrared
evolved as $\sim (1 + z)^Q$ with $Q \sim 3$. This is consistent with 
observations of galaxy counts using {\it ISOCAM}. On the other hand, the 
analysis of
{\it COBE-DIRBE} observations favors a stronger evolution in far-infrared
emissivity with $Q > 4$. 

There are various possible interpretations of these results; among them is
the possibility that special relativity is modified at high energies.
Future $\gamma$-ray and infrared observations will be needed to resolve
this situation.

\acknowledgments
I would like to acknowledge my collaborators in much of this work. They are
(in alphabetical order) Okkie De Jager, Sheldon Glashow, Matthew Malkan and 
Michael Salamon.

\end{document}